        \documentstyle[11pt]{article}
\def\lsim{\, \lower2truept\hbox{${<
\atop\hbox{\raise4truept\hbox{$\sim$}}}$}\,}
\def\gsim{\, \lower2truept\hbox{${>
\atop\hbox{\raise4truept\hbox{$\sim$}}}$}\,}
\def\mincir{\ \raise -2.truept\hbox{\rlap{\hbox{$\sim$}}\raise5.truept  
\hbox{$<$}\ }}                                                          %
\def\magcir{\ \raise -2.truept\hbox{\rlap{\hbox{$\sim$}}\raise5.truept  %
\hbox{$>$}\ }}                                                          %
\def\frac#1/#2{\leavevmode\kern.1em
\raise.5ex\hbox{\the\scriptfont0 #1}\kern-.1em
/\kern-.15em\lower.25ex\hbox{\the\scriptfont0 #2}}

\font\piccolo=cmr8

\begin{document}

\begin{center}

{\piccolo Capri CMB Workshop Proceedings}

\smallskip
{\piccolo Astrophys. Lett. and Comm., in press}

\bigskip\bigskip\bigskip

{\Large FOREGROUND CONTRIBUTIONS TO 0.2-2$^{\circ}$ CMB ANISOTROPIES}

\bigskip\bigskip

L. TOFFOLATTI\footnote{Osservatorio Astronomico di Padova, Italy;
$^{2}$~Dipartimento di Astronomia di Padova, Italy; $^{3}$~ITESRE -- CNR,
Bologna, Italy; $^{4}$~Lawrence Berkeley Laboratory, Berkeley, CA, USA;
$^{5}$~IFCTR -- CNR, Milano, Italy; $^{6}$~Dipartimento di Fisica,
Universit\`a di Tor Vergata, Roma, Italy; $^{7}$~University of Cambridge,
Cambridge, UK; $^{8}$~Haverford College, Haverford, USA; $^{9}$~NRAL,
Jodrell Bank, UK; $^{10}$~Dipartimento di Fisica, Universit\`a di Milano,
Milano, Italy; $^{11}$~CEN/CEA -- SAP, Saclay, France; $^{12}$~Universidad
de Cantabria, Santander, Spain; $^{13}$~Istituto de Astrofisica de Las
Canarias, Tenerife, Spain; $^{14}$~Universidad de Valencia, Valencia, Spain;
$^{15}$~Dipartimento di Fisica, Universit\`a La Sapienza, Roma, Italy;
$^{16}$~European Southern Observatory, Garching, Germany;
$^{17}$~Jet Propulsion Laboratory, Pasadena, CA, USA; $^{18}$~IAS,
Universit\'e de Paris XI, Orsay, France; $^{19}$~Istituto Universitario
Navale, Napoli, Italy; $^{20}$~Max Planck Institut f\"ur Physik, M\"unchen,
Germany; $^{21}$~Observatoire de Meudon, Meudon, France; $^{22}$~Osservatorio
Astronomico di Arcetri, CAISMI--CNR, Arcetri, Italy; $^{23}$~Osservatorio
Astronomico di Capodimonte, Napoli, Italy; $^{24}$~Osservatorio Astronomico
di Roma, Monte Porzio, Roma, Italy; $^{25}$~University of Oxford, Oxford, UK},
L. DANESE~$^{2}$, A. FRANCESCHINI~$^{1}$, N. MANDOLESI~$^{3}$,
G.F. SMOOT~$^{4}$, M. BERSANELLI~$^{5}$, N. VITTORIO~$^{6}$, A. LASENBY~$^{7}$,
R.B. PARTRIDGE~$^{8}$, R. DAVIES~$^{9}$, G. SIRONI~$^{10}$,
C. CESARSKY~$^{11}$, M. LACHIEZE--REY~$^{11}$, E. MARTINEZ--GONZALEZ~$^{12}$,
J. BECKMAN~$^{13}$, R. REBOLO~$^{13}$, D. SAEZ~$^{14}$,
P. DE BERNARDIS~$^{15}$, G. DALL'OGLIO~$^{15}$,
P. CRANE~$^{16}$, M. JANSSEN~$^{17}$, J.L. PUGET~$^{18}$,
E. BUSSOLETTI~$^{19}$, G. RAFFELT~$^{20}$, P. ENCRENAZ~$^{21}$,
V. NATALE~$^{22}$, G. TOFANI~$^{22}$, P. MERLUZZI~$^{23}$,
R. SCARAMELLA~$^{24}$, AND G. EFSTATHIOU~$^{25}$

\end{center}
\bigskip\bigskip\bigskip

{\piccolo
Send proofs to: \hskip 3truecm Luigi Toffolatti

\hskip 5.1truecm Osservatorio Astronomico di Padova

\hskip 5.1truecm vicolo dell'Osservatorio, 5\ \ I-35122 PADOVA (Italy)

\hskip 5.1truecm FAX: +39--49--8759840

\hskip 5.1truecm e-mail: 39003::TOFFOLATTI (SPAN)

\hskip 5.1truecm or TOFFOLATTI@ASTRPD.PD.ASTRO.IT}

\bigskip\bigskip
{\bf Abstract:} We examine the extent to which galactic and
extragalactic foregrounds can hamper the detection of
primordial Cosmic Microwave Background (CMB) anisotropies.
We limit our discussion to intermediate angular scales,
$10^{\prime}\lsim \theta \lsim 2^{\circ}$, since
many current as well as future experiments have been designed
to map CMB anisotropies at these angular scales.
In fact, scales of $\gsim 10^{\prime}$ are
of crucial importance to test both the conditions in the early Universe
and current theories of the gravitational collapse.

\newpage
\baselineskip=22pt
\bigskip
\section{INTRODUCTION}

Our purpose here is to re-estimate the contributions of
the Galaxy and of discrete extragalactic sources to the Cosmic Microwave
Background (CMB) fluctuations, focusing on intermediate angular scales,
$10^{\prime}\lsim \theta \lsim 2^{\circ}$. These angular scales
are soon becoming the most interesting ones to confirm the $COBE$ DMR
detection of large--scale primordial anisotropies (Smoot et al. 1992;
Wright et al. 1992). In fact,
scales of order 0.2-2$^{\circ}$ provide crucial information on the nature
of dark matter, the existence of topological defects, the thermal
history of the Universe and the imprints left on the CMB by gravitational
effects after the recombination. Thus sensitive observations of the CMB
anisotropy at $\lsim 2^{\circ}$ resolution are probably one of the most
important tools of observational cosmology.

Given the very strong interest on the subject, many ground--based and
balloon--borne as well as satellite experiments have recently been
developed or proposed to study CMB anisotropies at these angular scales.
The atmospheric emission is a serious problem for ground--based
experiments while it is largely alleviated in balloon ones. On the other
hand, galactic and extragalactic foregrounds are a major problem for any
kind of experiment aiming to push the uncertainties at the $\Delta T/T
\simeq 10^{-6}$ level. Therefore, a thorough analysis of the foreground
emission, spanning the whole wavelength range from $\sim$ 1 cm down to 400-500
$\mu$m, is of great interest to identify the best spectral window where
foreground anisotropies reach their minimum value. As pointed out by Brandt
et al. (1994), multi--channel anisotropy measurements spanning two or three
octaves in frequency near the minimum region could perform very well in
distinguishing between truly primordial anisotropies and foreground ones.

The galactic radio continuum emission, which is the major source of the diffuse
background below $\sim$ 0.5-1 GHz, still gives the dominant contribution
to the foreground radiation up to $\nu \sim 100$ GHz, due to the combined
synchrotron and free-free emissions. At higher frequencies, beyond
the minimum at 80--120 GHz, emission from interstellar cold dust
starts to dominate and even around the intensity peak of the CMB
the Galaxy still yields a background relevant to our estimates
of the CMB intensity fluctuations. Anyway, galactic foreground fluctuations
can be estimated and possibly removed using multifrequency data,
providing that the spatial and spectral regions where this foreground
is large are avoided.

Concerning extragalactic radio sources, it is now possible,
thanks to the recent very deep VLA surveys of radio sources at cm wavelengths
(Windhorst et al. 1985; Condon \& Mitchell 1984; Fomalont et al.
1984; Partridge et al. 1986; Fomalont et al. 1988; Fomalont et al. 1991;
Windhorst et al. 1993) to derive essentially model independent estimates of
Poisson fluctuations down to scales $\simeq 30^{\prime\prime}$. Our predictions
are made uncertain only by the required extrapolation of the radio source
counts (which are directly measured at $\lambda\gsim 3$ cm)
to mm wavelengths.
There is, however, sufficient spectral information to substantially
reduce such an uncertainty (at least for the dominant source populations).

At $\nu\gsim$ 100 GHz, the contribution of far--IR selected sources, whose
emission is dominated by interstellar dust, starts to be appreciable.
Since our knowledge of the counts of extragalactic far--IR sources still
relies on the $60 \mu$m IRAS survey,
their contribution to CMB anisotropies is difficult to estimate due
to the lack of data at faint fluxes ($\lsim 50$ mJy) and to
the yet poor knowledge of cold dust components in galaxies.
Estimates of the $\Delta T/T$ levels due to Poisson distributed far--IR
sources has been calculated by Franceschini et al. (1989,1991) while Wang
(1991)
focused on the $\Delta T/T$ contribution given by a non--Poisson
distribution of far--IR galaxies.

Correlations in the spatial distribution of extragalactic radio sources, for
which there is increasing evidence, provide an additional signal to the
background anisotropies. The amplitude of the effect is difficult to
estimate, but preliminary indications of substantial clustering for radio
galaxies of medium radio power (Peacock \& Nicholson, 1991) suggest
that the non--Poisson contribution to fluctuations may be important on some
angular scales. As for far--IR/sub--mm sources, the information on their
clustering properties still relies on the IRAS survey at 60 and 100 $\mu$m.
Since dust emission from galaxies gives a great contribution to the
far--IR/sub--mm background, a non--Poisson distribution of sources
could give non negligible contributions to CMB anisotropies at sub--mm
wavelengths. Anyway, analyses of the IRAS data have shown that the correlation
length of disk galaxies is smaller than that of optically or radio selected
galaxies, while there is no up--to--now evidence of correlations for
early--type IRAS galaxies. Thus, the $\Delta T/T$ contribution
due to clustered sources in the far--IR/sub--mm wavelength range is
smaller than in the radio and we will only briefly comment on this point.

\section{ANISOTROPIES DUE TO GALACTIC EMISSION}

At present, full--sky radio maps at resolutions comparable to
(or better than) those discussed here are only available
at decimeter wavelengths (Haslam et al., 1982; Reich \& Reich, 1982).
Unfortunately, they are rather poorly calibrated, with a 5--10\%
typical uncertainty on temperature variations. The situation is better
in the far--IR/sub--mm domain, where the properly
calibrated 100 $\mu$m IRAS and 240 $\mu$m $COBE$ DIRBE maps
are well suited to predict galactic anisotropies. In this case,
the problem resides in the required large spectral extrapolation of
energy distributions steeply varing with the wavelength.

A comprehensive analysis of the confusion noise given by galactic
synchrotron radiation (GSR) and dust emission has been recently performed by
Banday \& Wolfendale (1990; 1991a,b) and by Banday et al.
(1991). Their main conclusions are the following. {\it a)} The GSR
fluctuations at frequencies higher than $\sim$ 20 GHz should be small enough
($\Delta T/T\lsim 5\times 10^{-6}$) not to dominate the cosmological effect.
Therefore, with some improvements in the GSR noise predictions and in the
technical quality of the observations, the
adoption of frequencies $\gsim$ 30 GHz would allow to detect true
CMB anisotropies. {\it b)} As for the galactic dust, they took
into account the nature of the dust emission, in terms of grain
properties and environment and discussed the overall Galactic emission,
focusing on high galactic latitudes where the dust is heated by
the general interstellar radiation field. Considering GSR and galactic dust
noise together, they found that the lowest $\Delta T/T$ achievable, away from
currently known cirrus complexes, is $\sim (2-4)\times 10^{-6}$ at $\sim$ 90
GHz on the angular scales of the $COBE$ DMR experiment.

Another estimate of the $\Delta T/T$ levels due to the Galaxy has been given
by Masi et al. (1991). They estimated the general spectrum of the diffuse
galactic emission from available experimental data. Moreover,
by a pixel--to--pixel correlation between the 408 MHz
map (Haslam et al., 1982) and the IRAS 100 $\mu$m emission
and avoiding low galactic latitude regions ($\vert b\vert > 5^{\circ}$)
they found a very significant spatial correlation between the cm radio and the
far--IR emission in our Galaxy. Their main result is that there is a
spectral window ($0.5 > \lambda > 0.11$ cm) and many spatial windows (5--10$\%$
of the sky) where anisotropies due to the galactic emission keep below
$\Delta T/T \sim 2-4\times 10^{-6}$, on angular scales $0.5^{\circ}
\lsim \theta \lsim 5^{\circ}$.

Concerning the free-free emission, the only full--sky maps
available up to now at frequencies where free--free should dominate
are those provided by $COBE$ DMR. The analyses of Bennett
et al. (1992, 1994) of the DMR data show that their ``pure free--free'' map at
53 GHz has a galactic latitude dependence of $T^{ff}_{A}(\mu K)= (10\pm 4)
{\rm cosec}\vert b\vert$ for $\vert b\vert > 15^{\circ}$. On the other hand,
probes of the warm ($T\approx 10^4$ K), low density ($n\approx 10^{-1}$
cm$^{-3}$ ionized medium such as $H_{\alpha}$, $C^+$ and $N^+$ are relevant
indicators of the hydrogen free--free continuum at intermediate to high
galactic latitudes (Reynolds, 1992; Bennett et al., 1992, 1994) and
the associated free--free emission at radio frequencies can then be directly
calculated from the observed $H_{\alpha}$ intensity.

Using a $\sim 0.8$ deg beam, Reynolds (1992) found that the average free--free
intensity at high galactic latitudes predicted by the $H_{\alpha}$ background
is
$1.2\times 10^{-6}$ and $6\times 10^{-6}$ times that of the CMB at
$\lambda$=3.3 and 9.5 mm, respectively. The same author also derived
an amplitude of the free--free cosecant law a factor of $\sim$3 smaller
than that obtained from the $COBE$ full sky maps. This can be understood
by considering that Reynolds picks up location that are free from discrete
sources, while $COBE$ does not make any source exclusion
(see Bennett et al., 1992). Anyway, since bright spots could be
identified and subtracted with high resolution observations, this could
be also interpreted in terms of a smaller free--free contribution to CMB
fluctuations at intermediate to small angular scales ($\lsim 1^{\circ}$).

\subsection{$\underline{\rm Sky\ Fluctuations\ at\ High\ Galactic\
Latitude}$}

The Gautier III et al. (1992) spatial power spectrum analysis
of the sky surface brightness provided by the IRAS
100 $\mu$m maps and scans is a useful tool to estimate the confusion noise due
to infrared cirrus. By this formalism, and fixing the average sky brightness
at 100 $\mu$m, one can easily calculate the $\Delta T/T$ levels at
different angular scales and for different configurations of the observing
reference aperture. Then, adopting a suitable emission spectrum, it is possible
to extrapolate the estimated $\Delta T/T$ levels to longer wavelengths (see
also Franceschini et al., 1991) under the assumption that the different
emission components in our Galaxy are {\it spatially correlated} at different
angular scales. This correlation is clearly associated with global properties
of the ISM in the galactic disk (i.e. energy balance among the different
processes) whereas it cannot be directly translated to every single observed
region of the Galaxy without a critical discussion. There are likely local
changes in the synchrotron emission due to fluctations in the distribution
of the ISM from supernova shocks, winds from OB stars and variations in
the irregular component of the galactic magnetic field (Banday et al., 1991;
Bennett et al., 1992). At the same time, there is some evidence of a more
diffuse distribution for the free--free component (Hancock et al., 1994).
Anyway, the diffuse ionized hydrogen appears correlated with the dust emission
at high galactic latitude not only as regards the diffuse emission but also
in local patches of the sky
(HII regions have been clearly identified associated with O and B stars and
bursts of star formation). As for the synchrotron component, it is very
strongly correlated with the dust emission as proved by radio and IRAS
observations of disk galaxies (i.e. de Jong et al., 1985; Helou et al., 1985)
and, for our Galaxy, by the analysis of Masi et al. (1991), who found a strong
correlation on a pixel size of $2^{\circ}\times 2^{\circ}$, comparable with the
angular scales of interest here. Moreover, at high galactic latitude,
the column density of the HII is found to range between 26\% and 63\% of that
of the HI (Reynolds, 1991) while this latter shows a well known correlation
with the 100 $\mu$m dust emission (Boulanger \& Perault, 1988).
For all these reasons, we think that our assumption is likely to
give at least a first--order estimate of the anisotropies due to the galactic
foreground.

To fix the average sky brightness, $B_0$, at high galactic latitude
we consider that 10$\%$ of the sky has $B_0\leq 1.5$ MJy/sr at 100 $\mu$m.
Furthermore, as reported by  Lockman et al.\ (1986)
about 8$\%$ of the sky has HI column densities $N_H\lsim 1.5
\times 10^{20}\ cm^{-2}$. Using the correlation between 100 $\mu$m and
HI emission derived by Boulanger \& Perault (1988) such column densities would
imply a sky brightness $<$1.3 MJy/sr. Considering the capabilities
of future multi--channel high sensitivity experiments in the sub--mm domain,
which should allow to subtract -- at least partially -- the dust emission,
we have adopted a 100 $\mu$m average residual sky brightness
which is 0.4 MJy/sr, $\sim$30$\%$ of the previously quoted value.

To estimate the confusion noise in the sub--mm
domain we adopted the model of Rowan--Robinson (1992) to describe
the far--IR/sub--mm spectrum of the Galaxy.
By incorporating very large grains to explain excess
emission at millimetre wavelengths, the model presented by Rowan--Robinson
provides an excellent fit to the interstellar extinction curve and
to the far--IR spectrum of dust in our Galaxy. In particular, we used
his fit to the emissivity towards the galactic pole (Rowan--Robinson, 1992,
Figure 3a) to extrapolate down in frequency the $\Delta T/T$
levels calculated at 100 $\mu$m. The resulting dust spectrum presents a
frequency dependence $\nu^{\alpha}B(\nu)$ with ${\alpha}=1.5$--1.7 and
a steepening to ${\alpha}\simeq 2$ at lower frequencies ($\nu\lsim$400--500
GHz).

As regards the estimated galactic noise in the radio, at $\lambda \gsim$3 mm,
we have avoided to use the radio maps at lower frequencies due to their
poor calibration and for they suffer from both striation and baseline
problems which hinder the prediction of anisotropy levels. We assumed that
the very tight correlation between the radio centimetric and far--IR/sub--mm
emissions for the galactic disc component (de Jong et al., 1985 and references
therein; Rowan--Robinson, 1992) holds also at high galactic
latitude at intermediate angular scales ($\sim 1^{\circ}$) which seems
to be confirmed by the analysis of Masi et al. (1991).
To extrapolate our predictions from
1.4--5 GHz up to 90--100 GHz we have then adopted the spectral
indices of the free--free and synchrotron emissions estimated by Bennett et al.
(1992), assuming that synchrotron and free--free give and equal contribution
to the total galactic emission at $\simeq$ 30 GHz and $B=1.4\mu$G for the
magnetic field strength.

In spite of the uncertainties, due to the assumed angular and spectral
dependence, our estimates of the galactic
noise agree with previously published results based on quite different and
independent assumptions (see \S 4.1).

\section{ANISOTROPIES DUE TO EXTRAGALACTIC SOURCES}

\subsection{$\underline{\rm Sky\ Fluctuations\ from\ Randomly\ Distributed\
and\ Clustered\ Sources}$}

Since the problem has been extensively discussed in the literature, we will
only sketch it here. All our estimates are based on the assumption of
``point--like'' sources (Rowan--Robinson \& Fabian, 1974): because the
angular scales of interest here are larger than the typical source size, we
can be confident that this assumption should not
affect our predictions. We defer to Franceschini et al. (1989) and references
therein for a thorough discussion on fluctuations from randomly distributed
sources.

The contribution to the intensity fluctuations from clustered source
populations can be computed under rather general hypotheses. The formalism
adopted has been developed and discussed by De Zotti et al. (1990, 1994) and
Mart\'\i n-Mirones et al. (1991) and we will defer to these articles for all
the relevant formulae.

While Poisson noise is dominated by sources counts at fluxes corresponding to
$\sim 1$ source per beam, hence by sources in a {\it limited flux interval}
(the most abundant fainter ones producing only smaller contributions),
conversely, {\it all fluxes} contribute to non-Poisson noise, which is then
dominated by {\it the faintest and more numerous sources which do actually
cluster} (see Barcons, 1992 for more details).

Pushing the detection threshold to fainter and fainter limits will then
reduce the Poisson contribution to the sky fluctuations at a fixed angular
scale more than the additional one due to the clustering. So, if {\it all
sources} do cluster, even with a relatively small clustering length, $r_0$,
we can have a $(\Delta T/T)_{cl}$ level of the same order of the Poisson one.

\subsection{$\underline{\rm Radio\ Source\ Counts\ and\ their\ Extrapolation\
to\ high\ Frequencies}$}

Our predictions of the expected fluctuation levels due to discrete
sources are based on the interpretation of deep survey data
at cm wavelengths proposed by Danese et al. (1987).
Adopting a simple evolutionary scheme, they have been able to explain the
flattening of the 1.4 and 5 GHz source counts, the identification statistics
and redshift distributions at sub-mJy flux levels, as well as data at bright
flux densities.

All available analyses (e.g. Impey \& Neugebauer 1988) have clearly shown
that compact sources, which dominate the source counts at wavelengths
$\lambda\leq 1$ cm, have spectral indices keeping ``flat'' ($\alpha \simeq
0.0$, although with some scatter) at least up to $\sim 100$ GHz, bending down
only at $\sim 10^{12}$ Hz. A further test on the high frequency behaviour of
compact radio sources can be obtained by comparing predictions on source
counts with data from high frequency complete surveys. Franceschini et al.
(1989) have compared 10 GHz source counts with model predictions
extrapolated to this frequency with spectral indices of flat-spectrum sources
above 2.4 GHz allowed to vary from 0 to 0.4: they found
that the 10 GHz data are consistent with an average spectral index $\sim 0$
for compact sources.

In view of the above, we have considered, for compact sources,
the case of $\alpha =0$ over the whole frequency range of
interest here. Above $\nu =100$ GHz we have assumed that spectra break to
$\alpha =0.7$. For the steep spectrum sources, we have adopted the radio
power -- spectral index relation determined by Peacock and Gull (1981).

\subsection{$\underline{\rm The\ Contribution\ of\ Far-IR\ Sources}$}

Far--IR selected sources, whose emission is dominated by the cold and warm
dust components, are likely to originate high CMB fluctuations in the sub--mm
domain, due to their rapidly increasing integrated intensity, $\nu I(\nu)$,
if compared to the fast decrease of the CMB shortwards of the peak at $\sim$
1 mm. At wavelengths longer than the CMB intensity peak, far--IR/sub-mm
sources still give a non-negligible contribution, at least down to $\sim
80-90$ GHz, depending on the adopted far--IR to sub--mm spectral behaviour
and on the angular scale. Extrapolations to still longer wavelengths
give negligible contributions to the predicted
$\Delta T/T$ levels for any reasonable thermal dust spectrum.

We adopt here two different descriptions of the cosmological evolution of
far--IR sources, in order to provide a confidence interval to the predicted
$\Delta T/T$ level. The first one is the model by Franceschini et al.
(1988), assuming a strong cosmological evolution of galaxies
with significant star formation activity (Actively Star Forming galaxies),
while the second one predicts appreciable cosmological evolution of both
late- and early-type galaxies (see Franceschini et al. 1994 for more
details).

Apart from ellipticals and S0s, for which there is virtually no information,
the spectra of all other galaxy populations in the far--IR and sub--mm domains
are dominated by thermal dust emission. The first model
adopts the far--IR/sub--mm spectrum of Kreysa \& Chini (1989), which is
based on a very high $(f_{\rm 1.25\ mm}/f_{\rm 100\mu m})\sim
1.5\times 10^{-2}$ flux ratio for normal galaxies, 3-5 times higher than
found by Andreani \& Franceschini (1992). The spectra adopted in the
second, more conservative, model
have been derived from the photometric model by Xu \& De Zotti (1989)
for Spiral and ASF galaxies, which is based on available IRAS and
sub--mm data. These spectra turn out to be in good agreement with recent
millimeter observations of a complete sample
of $IRAS$ selected galaxies (Andreani \& Franceschini, 1992).
The far--IR spectrum of Seyfert galaxies has been derived
from observations of AGNs by Chini et al. (1989) and Ward et al. (1987)
(see Franceschini et al. 1991, for more details).

The first model, to be taken as a somewhat extreme limiting case, gives the
highest integrated background
due to sources still compatible with current FIRAS upper limits on the
extragalactic component (see Wright et al. 1993). We believe that
the second one, based on a broad--band spectral description for
galaxies which takes into account in detail the evolution of the
thermal dust emission with cosmic time (Mazzei et al. 1992, 1994), should
more realistically predict the average sky noise due to galaxies.


\section{RESULTS AND DISCUSSION}

We summarize in Figure 1 our predictions on confusion noise due either to
galactic emission as well as to extragalactic discrete sources.
The three panels refer to three angular scales ($2^{\circ}$, $0.5^{\circ}$,
and $10^{\prime}$) covering the most relevant range for current and future
experiments on CMB anisotropies. As reminded in \S 1, these
angular scales provide informations on primordial density fluctuations on
scales corresponding to the observed large--scale structures, allowing
to study the physics of structure formation.
The figure covers the whole frequency range (20-500 GHz) around
the CMB intensity peak to identify the
most suitable frequencies for future experiments.

\subsection{$\underline{\rm Galactic\ Noise}$}

Our current estimate, which relies critically on the Gautier III et al. (1992)
analysis, and in particular on their claimed dependence of the
galactic noise on the angular scale ($\Delta I/I \propto \theta^{0.45}$),
shows that the Galaxy and the extragalactic
sources give comparable contributions to the CMB anisotropies for
$\theta\simeq 0.5$ deg, while at smaller scales extragalactic sources
dominate. At scales $\gsim 1^{\circ}$ the galactic emission gives by far
the dominant contribution to the confusion noise (upper panel of Figure 1).

We also confirm that the lowest $\Delta T/T$ levels due to
galactic synchrotron, free-free and dust radiation are found around $\nu \sim$
100 GHz. In particular, our best guess at 90 GHz results in good agreement
with that obtained by Banday \& Wolfendale (1991)
by convolving the IRAS HCON2 maps with the $COBE$ beam size of $7^{\circ}$ and
assuming the Lubin's et al. (1990) dust emission spectrum
(if we take into account the different angular scales
examined in the two cases). A similar agreement is also found with the
Bennett et al. (1992) $COBE$ DMR estimate of the 90 GHz anisotropy level
due to high galactic latitudes dust, giving $\Delta T_A \simeq$ 2.3--6.5
$\mu$K.
Albeit uncertain within a factor $\sim$ 2-3, due to the meagre information
available on the spatial distribution and correlations of the weak high
latitude emission of the Galaxy, our current estimates indicate that the
galactic confusion noise should not be greater than $\Delta T/T\sim 10^{-6}$,
at least in the best spectral window.


\subsection{$\underline{\rm Extragalactic\ Sources}$}

\subsubsection{$\underline{\rm Poisson\ fluctuations}$}

Concerning the contribution of randomly di\-stri\-but\-ed sour\-ces,
it is clear that the choice of smaller scales does not help in
reducing their confusion noise. As already discussed by Franceschini et al.
(1989; 1991), the shape of the radio and far--IR source counts is such
that the highest $\Delta T/T$ levels due to extragalactic source
populations are reached at intermediate to small angular scales
($\theta \lsim 10^{\prime}$). For this reason, if radio and far--IR selected
sources give a negligible contribution to CMB fluctuations on the
large angular scale of the COBE DMR experiment, at smaller scales
they turn out to be increasingly important and for angular scales $\lsim
20^{\prime}$ they are probably the dominant foreground noise.

In particular, we obtain $(\Delta T/T)\simeq 4-5\times 10^{-6}$
at $\sim 30$ GHz, summing up the contributions of all the undetected sources
up to the $5\sigma$ detection limit: a value
very close to the level expected from
scale--invariant primordial fluctuations. At higher frequencies, $\sim$
50-90 GHz, we have a better situation, with $\Delta T/T \sim 1-2\times
10^{-6}$, i.e. sufficiently small to only marginally affect the detection of
intrinsic CMB anisotropies.
This decrease of the predicted fluctuation levels at higher frequencies
is due both to the
fading of steep--spectrum radio sources, which give a progressively
smaller contribution to the counts, and to the ordinary K--correction
in high-redshift bright ``flat''--sources observed beyond
the steepening of the spectrum at 100 GHz.

Far--IR selected sources start to contribute to Poisson
CMB fluctuations at, say, 80-100 GHz, strongly depending on the assumed
dust spectrum, as discussed in \S 3.3. For any reasonable
choice of the spectrum, they do not contribute more than $10-20\%$ to the
total predicted fluctuations at $\theta = 30$ arcmin. On the other hand, they
start to dominate at smaller angular scales ($\theta \lsim 5-10$ arcmin),
with a contribution at $\theta =1^{\prime}$ a factor $\sim 5$ higher
than that coming from radio selected sources.
The very steep dust emission spectra and the few data available in the
sub--mm region on the emission of extragalactic sources suggest to avoid
wavelengths shorter than $\lambda\simeq 1.2-1.3$ mm to search for intrinsic
CMB anisotropies.
Indeed, Poisson fluctuations rapidly increase for $\lambda <1.5$ mm:
at $\lambda\simeq 1$ mm their level, although uncertain to within a
factor of $\sim$ 3 because of the different choices of the cold-dust spectrum
and of the cosmological evolution of sources, is again well above
$(\Delta T/T)\simeq 10^{-6}$ at $\theta\simeq 10^{\prime}$, while keeping
just below the $10^{-6}$ level for $\theta =30^{\prime}$.


\subsubsection{$\underline{\rm Non-Poisson\ fluctuations}$}
Our estimates of the non-Poisson contribution to $\Delta T/T$
are all based on the analysis of large scale
clustering of radio galaxies done by Peacock \& Nicholson (1991). They
have found that radio source cluster, at least for a particular power range
(log$P_{\rm 1.4\ GHz}\simeq 22.5-24.5 h^{-2}$ WHz$^{-1}$sr$^{-1}$).
Since we have assumed that all galaxies contributing to the background
anisotropies do cluster, our estimates could be taken as a safe upper limit.
On the contrary, if we consider that only sources in the power ranges
for which clustering has been detected do actually cluster, then the
predicted contribution of clustering to the fluctuation level will be
significantly reduced for the same choice of the other relevant parameters.


What is found is that the non-Poisson contribution to the anisotropies is
usually negligible, being a factor of $\sim 2.5-3$ below the Poisson noise
even adopting the highest reasonable value for the clustering length.
Owing to the different dependence on the assumed flux limit (see \S 3.1),
only for experiments in which source identification and subtraction is
performed down to flux levels much fainter than the 5$\sigma$ detection
threshold and assuming the same clustering length as found by Peacock \&
Nicholson for all source luminosities, would the fluctuations due to
clustered sources be important at the angular scales of interest here.

Concerning sub--mm sources, no information is available on their clustering
properties. Anyway, since far--IR sources selected from the $IRAS$ Point
Source Catalogue seem to be characterized by a very small clustering length,
3-4 times smaller than found for radio sources, we can be confident that
their non-Poisson contribution to the CMB fluctuations keeps always
negligible.

\medskip\medskip\medskip

\leftline{\bf Acknowledgements} We wish to thank the referee, A.Banday, for
many useful comments and suggestions which helped us to improve the manuscript.
L.T. would like to thank the Universities of Oviedo and Cantabria (Spain)
for their hospitality during part of the preparation of this paper.
This work has been partially supported by the Commission of the European
Communities, ``Human Capital and Mobility Programme'' of the EC, contract
number CHRX--CT92--0033 and by the Agenzia Spaziale Italiana (ASI).

\newpage

\medskip\medskip\medskip

\centerline{\bf References}

\medskip\medskip\medskip

\def\ref{\noindent\hangindent=20pt\hangafter=1}

\ref
Andreani, P., \& Franceschini, A. 1992, A\&A, 260, 89

\ref
Banday, A.J., \& Wolfendale, A.W. 1990, MNRAS, 245, 182

\ref
Banday, A.J., \& Wolfendale, A.W. 1991a, MNRAS, 248, 705

\ref
Banday, A.J., \& Wolfendale, A.W. 1991b, MNRAS, 252, 462

\ref
Banday, A.J., Giler, M., Szabelska, B., Szabelski, J., \& Wolfendale, A.W.
1991, ApJ, 375, 432

\ref
Barcons, X. 1992, ApJ, 396, 460

\ref
Bennett, C.L., et al. 1992, ApJ, 396, L7

\ref
Bennett, C.L., et al. 1994, ApJ, in press

\ref
Bennett, C.L., Hinshaw, G., Banday, A., Kogut, A., Wright, E.L.,
Loewenstein, K., \& Cheng, E.S. 1993, ApJ, 414, L77

\ref
Boulanger, F., \& Perault, M. 1988, ApJ, 330, 964

\ref
Brandt, W.N., Lawrence, C.N., Readhead, A.C.S., Pakianathan, J.N., \&
Fiola, T.M. 1994, ApJ, in press


\ref
Chini, R., Krugel, E., Kreysa, E., Gemund, H.-P. 1989, A\&A, 216, L5


\ref
Condon, J.J., Mitchell, K.J. 1984, AJ, 89, 610


\ref
Danese, L., De Zotti, G., Franceschini, A., \& Toffolatti, L. 1987, ApJ,
318, L15

\ref
de Jong, T., Klein U., Wielebinski R., Wunderlich, E. 1985, A\&A, 147, L6

\ref
De Zotti, G., Persic, M., Franceschini, A., Danese, L., Palumbo, G.G.C.,
Boldt, E.A., \& Marshall, F.E. 1990, ApJ, 351, 22

\ref
De Zotti, G., Franceschini, A., Toffolatti, L., \& Mazzei, P., 1994,
Astrophys. Lett \& Comm., submitted


\ref
Fomalont, E.B., Kellermann, K.I., \& Wall, J.V. 1984, ApJ, 277, L23

\ref
Fomalont, E.B., Kellermann, K.I., Anderson, M.C., Weistrop, D.,
Wall, J.V., Windhorst, R.A., \& Kristian, J.A. 1988, AJ, 96, 1187

\ref
Fomalont, E.B., Windhorst, R.A., Kristian, J.A., \& Kellermann, K.I.
1991, AJ, 102, 1258

\ref
Franceschini, A., Danese, L., De Zotti, G., \& Toffolatti, L. 1988,
MNRAS, 233, 157

\ref
Franceschini, A., Toffolatti, L., Danese, L., \& De Zotti, G. 1989,
ApJ, 344, 35

\ref
Franceschini, A., Toffolatti, L., Mazzei, P., Danese, L., \& De~Zotti, G.
1991, A\&A Suppl. 89, 285

\ref
Franceschini, A., Mazzei, P., Danese, L., \& De~Zotti, G.
1994, ApJ, in press

\ref
Gautier III, T.N., Boulanger, F., Perault, M., \& Puget, J.L. 1992,
AJ, 103, 1313

\ref
Hancock, S. et al., 1994, Nature, 367, 333

\ref
Haslam, C.G.T., et al., 1982, A\&AS, 47, 1

\ref
Helou, G., Soifer, B.T., Rowan-Robinson, M. 1985, ApJ, 298, L7

\ref
Impey, C.D., \& Neugebauer, G. 1988, AJ, 95, 307




\ref
Kreysa, E. \& Chini, R. 1989, Proc. {\it 3rd ESO/CERN Symp. on Astronomy,
Cosmology and Fundamental Physics}, eds. Caffo, M., Fanti, R., Giacomelli,
G., Renzini, A.. Kluwer, Dordrecht, p.433.


\ref
Lockman, F.J., Jahoda, K., \& McCammon, D. 1986, ApJ, 302, 432

\ref
Lubin, P., Meinhold, P.R., \& Chingcuanco, A.O., 1990, The Cosmic
Microwave Background -- 25 Years Later, eds. Mandolesi, N. \&
Vittorio, N., p.115

\ref
Masi, S., de Bernardis, P., De Petris, M., Epifani, M., Gervasi, M., \&
Guarini, G. 1991, ApJ, 366, L51

\ref
Mart\'{\i}n-Mirones, J.-M., De Zotti, G., Boldt, E.A., Marshall, F.E.,
Danese, L., Franceschini, A., \& Persic, M. 1991, ApJ, 379, 507

\ref
Mather, J.C. et al. 1990, ApJ, 354, L37

\ref
Mather, J.C. et al. 1993, ApJ, in press

\ref
Mazzei, P., Xu, C., \& De Zotti, G. 1992, A\&A, 256, 45

\ref
Mazzei, P., De Zotti, G., \& Xu, C. 1994, ApJ, in press





\ref
Page, L.A., Cheng, E.S., \& Meyer, S.S. 1990, ApJ, 355, L1

\ref
Partridge, R.B., Hilldrup, K.C., \& Ratner, M.I. 1986, ApJ, 308, 46



\ref
Peacock, J.A., \& Gull, S.F. 1981, MNRAS, 196, 611


\ref
Peacock, J.A., \& Nicholson, S.F. 1991, MNRAS, 253, 307


\ref
Reich, P. \& Reich, W. 1986, A\&AS, 63, 205

\ref
Reynolds, R.J. 1991, in IAU Symp. 144, The Interstellar Disk-Halo Connection
in Galaxies, ed. H.Bloemen (Dordrecht:Kluwer), p.67

\ref
Reynolds, R.J. 1992, ApJ, 392, L35

\ref
Rowan--Robinson, M., 1992, MNRAS, 258, 787

\ref
Rowan-Robinson, M., \& Fabian, A.C. 1974, MNRAS, 167, 419



\ref
Smoot, G.F., et al. 1992, ApJ, 396, L1


\ref
Xu, C., \& De Zotti, G. 1989, A\&A, 225, 12

\ref
Wang, B. 1991, ApJ, 374, 465

\ref
Ward, M.J., Elvis, M., Fabbiano, G., Carleton, N.P., Willner, S.P.,
Lawrence, A. 1987, ApJ, 315, 74

\ref
Windhorst, R.A., Miley, G.K., Owen, F.N., Kron, R.G., \& Koo, D.C. 1985,
ApJ, 289, 494

\ref
Windhorst, R.A., Fomalont, E.B, Partridge, R.B., \& Lowenthal, J.D. 1993,
ApJ, 405, 498

\ref
Wright, E.L., et al. 1992, ApJ, 396, L13

\ref
Wright, E.L., et al. 1993, ApJ, in press

\newpage
\medskip
\centerline{\bf FIGURE CAPTION}

\medskip
\medskip\medskip
{\bf Figure 1} {Estimated fluctuation levels,
in terms of the thermodynamic temperature $\Delta T/T$, due to the galactic
polar emission and to the extragalactic sources. The three panels refer
to different angular scales. The three plotted curves
indicating the galactic anisotropy levels refer to the following choices
for the dust emission spectrum: the central one refer to the Rowan--Robinson
(1992) model (see text for more details); the upper one adopts a less steep
dust emissivity index ($\alpha\simeq 1$) useful to give an
upper limit to galactic anisotropies and  accounting for a possible
low--frequency excess at high galactic latitude (due to large dust grains
having an enhanced emissivity) like that seen from many groups close
to the galactic plane (i.e. Page, Cheng \& Meyer, 1990); the lower one
assumes a steeper dust emissivity index ($\alpha=2$) and an higher dust
temperature (T$\simeq$24 K).
The two curves indicating the Poisson noise levels due to
discrete extragalactic sources refer to a source detection limit $x_c=5\sigma$
and to different models for the evolution of the cold dust.
The higher $\Delta T/T$ level corresponds to the model by Franceschini et al.
(1988) assuming strong cosmological evolution of the most luminous
far--IR selected sources (ASF galaxies), while the lower one refers
to a moderate cosmological evolution of both late- and early-type galaxies
(Franceschini et al., 1994).
Both models give integrated intensities $I(\nu)$ still compatible with
the recent $COBE$ FIRAS upper limits on the CMB residuals in the sub--mm domain
(Mather et al., 1993; Wright et al. 1993) but the higher is close to infringe
the FIRAS limits. The four frequencies foreseen for the COBRAS experiment,
31.5, 53, 90 and 125 GHz, are also indicated by the dotted vertical lines.}

\end{document}